\documentclass[aps,prl,twocolumn,groupedaddress,showpacs]{revtex4-1}

\usepackage{graphicx}
\usepackage{dcolumn}   
\usepackage{bm}        
\usepackage{amssymb}   
\begin{document}
\title{ Direct Evidence for the 'All-in/All-out' Magnetic Structure in the Pyrochlore Iridates from $\mu$SR}


\author{Steven M. Disseler}
\email{steven.disseler@nist.gov}
\affiliation{Center for Neutron Research, National Institute for Standards and Technology, Gaithersburg Maryland 20899}
\affiliation{Department of Physics, Boston College, Chestnut Hill, Massachussets 02467}
\date{\today}

\begin{abstract}

In the pyrochlore iridates, \textit{R}$_{2}$Ir$_{2}$O$_{7}$ (\textit{R} = Lanthanide, Y), determination of the magnetic structure of the iridium moments remains an outstanding problem despite the role this is expected to play in the formation of novel band structures and topologies in these materials. In this work, a new analysis of the experimentally measured spontaneous muon spin precession frequency is presented which incorporates both probabilistic and \textit{ab-initio} modeling techniques to determine the ground state magnetic structure. It is shown that the experimentally observed results are consistent only with a magnetically ordered Ir$^{4+}$ sublattice with the so-called 'all-in/all-out' magnetic structure, and that the electronic state of the Ir$^{4+}$ is best described by the $J_{eff} = \frac{1}{2}$ model in several member compounds. Through this approach it is also demonstrated that such a simple structure is not likely to be present on the rare-earth sublattice which contain much larger localized moments.

\end{abstract}


\maketitle
	
Oxides of \textit{5d} transition metals such as iridium exist at the intersection of two prominent areas of modern quantum materials research which probe many-body electron correlations and relativistic spin-orbit interactions (SOI) in condensed matter systems\cite{Pesin, Kim214, WKReview}. The former of these is exemplified in the Mott-insulating state found in many \textit{3d} metal-oxides\cite{Imada}, while the later is largely responsible for the topological nature of the band structure in materials such as Bi$_{2}$Te$_{3}$\cite{TI_Review}. The extended nature of the Ir-\textit{5d} orbitals results in electron correlation effects of the same energy scale as the SOI such that neither may be treated perturbitavely, and may result in a variety of novel phenomena and/or topological states\cite{Jiang_HK_Model, WKReview, Wan, Yang_Ran}. 

The pyrochlore iridate compounds, \textit{R}$_{2}$Ir$_{2}$O$_{7}$ (\textit{R} = Lanthanide, Y) have drawn significant interest in this regard, as early experiments revealed a finite temperature metal-insulator and magnetic transition which may be suppressed by increasing the size of the \textit{R}-site species\cite{Yanagishima, Matsuhira} until a metallic state develops for \textit{R} = Pr. Near the cross-over from insulator to metal a variety of topological insulator or semi-metal states are predicted to emerge, the nature of which depends strongly on the presence and symmetry of long-range magnetic order of the $J_{eff} = \frac{1}{2}$ $Ir^{4+}$ moments\cite{Wan, WKReview}. For example, the existence of the Weyl semimetal state may be found when the magnetic structure is of the so-called 'all-in/'all-out' (AIAO) type which preserves the inversion symmetry of the pyrochlore lattice\cite{Wan, Yang_Ran, WK_PRB}. It is for this reason that much of the experimental work to date has focused on determining the configuration of the magnetic Ir$^{4+}$ moments in these compounds.

However, while several probes have collectively provided a great deal of information about the underlying magnetic structure it has yet to be uniquely identified. For example, resonant x-ray diffraction measurements of Eu$_{2}$Ir$_{2}$O$_{7}$ indicate that long-range magnetic order occurs below $T_{MI}$ and is consistent with a magnetic propagation vector \textit{k} = (0, 0, 0) structure such as the AIAO\cite{Sagayama}. On the other hand, no such order has been observed in powder neutron diffraction measurements but from experimental resolution limits an upper limit of 0.5 $\mu_{B}$/Ir has been ascertained for this type of order, with a much smaller limit placed on \textit{k} $\neq 0$ structures \cite{Disseler2012, Shapiro}. To date, the most conclusive evidence for commensurate long-range order in these systems has come from zero-field muon spin relaxation (ZF-$\mu$SR), where it has been demonstrated that the order resides on the Ir$^{4+}$ sublattice in a wide range of \textit{R}$_{2}$Ir$_{2}$O$_{7}$ compounds \cite{Disseler2012, DisselerNd227, Zhao, Nd227New}. To determine the magnetic structure from $\mu$SR results however, one must have detailed knowledge of both the size of the magnetic moments and the location of the muon stopping site within the unit cell, neither of which have not been well described previously for these compounds. 

To overcome these difficulties, a two-step analysis of these previous experimental results is performed here to determine both the magnitude of the Ir$^{4+}$ moments and possible muon stopping sites. Together this approach will allow for potential magnetic structures to be distinguished from one another and the correct ground state to be selected based on experimental results. This approach first incorporates a technique recently developed for $\mu$SR utilizing Bayes'€™ theorem which produces probabilistic information about the size of the magnetic moment based on the experimentally observed muon precession frequency\cite{Blundell2012}. The information obtained from this calculation is then used in conjunction with \textit{ab initio} density functional theory (DFT) calculations to determine in greater detail information about the muon stopping site in order to distinguish potential structures from one another and identify the correct magnetic structure. 

A detailed description of the $\mu$SR technique can be found elsewhere \cite{Yaouanc}, however it is important to note that In a ZF-$\mu$SR measurement of a magnetically ordered system, the spontaneous muon precession frequency is given by $\nu = \gamma_{\mu}/2\pi |\mathbf{B}(r)|$ , where $\gamma_{\mu}/2\pi$ = 135.5 MHz/T is the gyromagnetic ratio€ of the muon, and $|\mathbf{B}(r)|$ is the magnitude of the magnetic field at the muon stopping site generated by the ordered moments. In the case of insulating antiferromagnetic materials, such as Y$_{2}$Ir$_{2}$O$_{7}$ and Eu$_{2}$Ir$_{2}$O$_{7}$, only the dipolar fields generated by the static magnetic moments give significant contributions to this static local field\cite{Yaouanc, BlundellDipole}, and thus the total field at the muon stopping site is given by,

\begin{equation}\label{Eq1}
\mathbf{B}(r)=\sum_{i}\frac{\mu_{i}\mu_{0}}{4\pi r^{3}}\left [ 3\left ( \mathbf{ \hat{\mu_{i}}\cdot\hat{r}} \right ) \mathbf{\hat{r} - \hat{\mu_{i}}}\right ]
\end{equation} 

where $\mu_{i}$ is the magnetic dipole moment of the $i^{th}$ moment, and $r$ is the distance between the muon stopping site and $i^{th}$ moment such that $r = \left| r_{i} - r_{\mu} \right|$ \cite{BlundellDipole}. 

From previous measurements of the spontaneous precision frequency it is clear the magnetic order occurs via a second-order phase transition \cite{Disseler2012, DisselerNd227, Zhao} and thus the underlying ground state should be described by a single irreducible representation of the of the \textit{Fd-3m} space group assuming a single magnetic propagation vector \textit{k} = (0, 0, 0).  These representations are described following the notation used by Ref.\cite{Willis}, with each $\Gamma_{i}$ given by one or more basis vectors $\psi_{j}$. Shown in Fig. 1, the configurations under consideration in this work include then the non-coplanar AIAO or $\psi_{1}$ ($\Gamma_{3}$), the $\psi_{3}$ ($\Gamma_{5}$) in which the spins are rotated by $\pi$/2 from the $\psi_{1}$ configuration, and the coplanar $\psi_{6}$ ($\Gamma_{7}$). Each of these configurations have been previously discussed as potential ground state based on theoretical calculations\cite{Wan, WKReview}, or x-ray studies of similar pyrochlore systems\cite{OsmiumPyrochlore, Willis}.

The Bayes' theorem approach proposed by Blundell \textit{et al}\cite{Blundell2012} is first employed to obtain a probability distribution function (PDF) for obtaining a particular magnitude for the Ir$^{4+}$ ordered moments given one of the configurations described above. The magnetic field is first calculated for 10$^{4}$ randomly generated stopping sites with the condition that each site lie at a radial distance 0.8 $< r < $1.1 {\AA}  away the nearest oxygen species and $r >$ 1 {\AA}  from any Ir sites, similar to that used in previous studies of other oxide materials\cite{Steele,Steele2}. To ensure proper convergence, the magnetic field was calculated for sites located at the center cell of 5 x 5 x 5 cube of unit cells, such that the total magnetic field at any given stopping site is the sum of some 2000  magnetic moments. When normalized properly, this yields a PDF $f(\nu | \mu)$, or the probability that a precession frequency is $\nu$ is observed given a set of moments of size $\mu$. Following the procedure outlined in Ref. \cite{Blundell2012} this function is then inverted to give a PDF $g(\mu | \nu)$ describing the probability of obtaining a specific moment $\mu$ given a known precession frequency $\nu$. This is succinctly given by,
	 
\begin{equation}\label{Eq2}
g\left (  \mu|\nu \right ) = \frac{ \frac{1}{\mu} f\left( \nu|\mu \right)} {\int_{0}^{\mu_{max}} \frac{1}{\mu'} f\left( \nu|\mu' \right) d\mu'  }
\end{equation} 

where the maximum possible moment $\mu_{max}$, here chosen to be 2 $\mu_{B}$ to prevent distortions to $g(\mu| \nu)$ which may arise from inducing an arbitrary small cutoff for the Ir moments.

\begin{figure}
\centering
  \includegraphics[width = 3.5in]{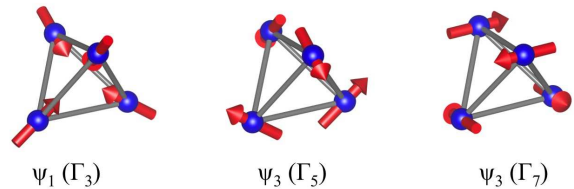}
   \caption{(Color Online) The three spin configurations of the Ir$^{4+}$ sublattice examined in this work. Each configuration is denoted by the single appropriate basis function following the notation in Ref. \cite{Willis} }
\label{Figure1}
\end{figure}

The $f(\nu | \mu)$ for each of the three spin configurations are shown in Fig. 2(a), using the structural information taken from the literature for Y$_{2}$Ir$_{2}$O$_{7}$\cite{Disseler2012}. Note that as the magnetic moment is assumed to carry a value of unity in the preceding calculations, the corresponding values of $\nu$ are given as $\nu/\mu$ and the subsequent units of $f(\nu | \mu)$  are such as to preserve the total probability under each curve. Each PDF is characterized by one or more maxima identifying the most probable values of the observed muon precession rate. However, as there are significant regions in which two or more spin-configurations have similar $f(\nu | \mu)$  additional information would be required to move beyond a probabilistic interpretation of these results such to distinguish clearly between spin-configurations. 

\begin{figure}
\centering
\includegraphics[width=3.5in]{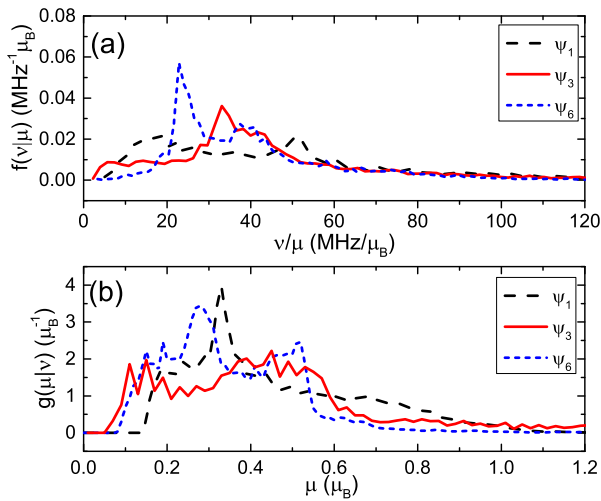}
\caption{(Color Online) (a) PDF for the frequency distribution $f(\nu | \mu)$  calculated for Y$_{2}$Ir$_{2}$O$_{7}$ by applying Eq. (1) to each of the spin configurations as described in the text. (b) Calculated probability distribution of the local  Ir$^{4+}$ moments $g(\mu|\nu)$ after application of Bayes' theorem for each of the configurations considered above.}
\label{Figure2}
\end{figure}

The $g(\mu| \nu)$ shown in Fig 2(b) have been calculated using the precession frequency observed in Ref. \cite{Disseler2012}. It is apparent that a significant portion of the probabilistic weight for all of these configurations is found below 0.5 $\mu_{B}$. This lies below the detection limit of the previous powder neutron diffraction measurements, corroborating these previous null-results. Examining in more detail, a single main peak is found for the $\psi_{1}$, while the $\psi_{3}$ and $\psi_{6}$ states have multiple broad peaks occurring on either side of this value, and extending up to the observable range ($\mu$ \textgreater   0.5  $\mu_{B}$). The location of the peak for $\psi_{1}$ˆ configuration at $\mu$  = 0.32 $\mu_{B}$ is nearly identical to that expected for an Ir$^{4+}$ ion with a $J_{eff}=\frac{1}{2}$ electronic configuration, where $\left< \mu \right> = gJ\mu_{B}$ = $\frac{1}{3}$ $\mu_{B}$. This result is consistent with the previous x-ray results\cite{Sagayama}, however as the other states do not have vanishing probability densities one cannot rule these out conclusively based on this analysis alone. 

\begin{figure}
\centering
\includegraphics[width=3.5in]{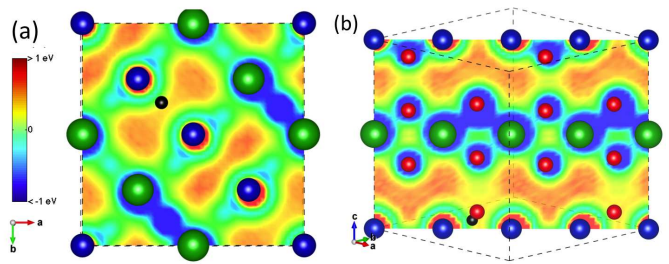}
\caption{(Color Online) Maps of the electrostatic potential inside the unit cell of Y$_{2}$Ir$_{2}$O$_{7}$ for the (a) the [001] plane located at c = 0.05\textit{a} and (b) [110] plane passing through the origin which intersect the muon stopping site, and visualized using VESTA \cite{Vesta}. Ir ions are shown in blue, Y  as green and O as red, and the muon as black.}
\label{Figure3}
\end{figure}

A more detailed comparison of these magnetic structures can also be obtained by determining the magnetic field at potential muon stopping sites. These have been investigated using DFT methods performed with the QUANTUM ESPRESSO\cite{QE} package within the generalized gradient approximation using ultrasoft psuedopotentials for the three atomic species. The muon was approximated by a norm-conserving, neutral hydrogen psuedopotential with the mass adjusted to that of the muon; this approach has been used to successfully determine the muon stopping site in numerous other material systems thus far\cite{QE_flourides, DFT_Muons}. A single unit cell of all 88 atoms plus the muon were used the calculation, with energy and charge density cutoffs taken to be 411 and 2466 eV respectively, and a 2 x 2 x 2 \textit{k}-point mesh used for Brilloiun zone integration. The muon was placed at over a dozen randomly generated interstitial sites and its position allowed to relax until the total energy of the structure reached a minimum self-consistent value. One site yielded a total energy more than 2 eV lower than any other and is indicated in Fig. \ref{Figure3}. Also shown relative to this site are the other atomic species and the calculated electrostatic potential for high-symmetry planes intersecting the muon stopping site. It can be see from this figure that this site is quite close to one of the 48\textit{f} oxygen sites as expected, and roughly bisects the nearest Ir-Ir bond. 

The muon precession frequency was then calculated and averaged over a volume of 0.01 $\AA^{3}$ centered at this site for each of the three test structures using the most probable value of the magnetic moment determined from the Bayes' theorem analysis, such that $\mu( \psi_{1} )$ = 0.32 $\mu_{B}$, $\mu(\psi_{3})$ = 0.24 $\mu_{B}$, and $\mu(\psi_{6})$ = 0.45 $\mu_{B}$.  The resulting frequencies are found to be 16.5 $\pm$ 0.1 MHz, 2.3 $\pm$ 0.3 MHz and 2.4 $\pm$ 0.3 MHz for the $\psi_{1}$, $\psi_{3}$, and $\psi_{6}$ configurations respectively. Despite the approximations made here, the field obtained for the $\psi_{1}$ configuration is remarkably close to the value measured in previous studies of Y$_{2}$Ir$_{2}$O$_{7}$ \cite{Disseler2012} that found $\nu$ = 14.9 $\pm$ 1.3 MHz. For either of the other structures examined here to yield this observed value would require a magnetic moment well over 1.5  $\mu_{B}$/Ir, well into the range observable by neutron scattering. Together with the results from the Bayes' theorem analysis these results provide strong evidence that the ground state configuration of the Ir lattice in Y$_{2}$Ir$_{2}$O$_{7}$ is $\psi_{1}$ with moment of approximately $\frac{1}{3}\mu_{B}$/Ir. 	 	

This technique can also be used to describe the magnetic structures of other pyrochlore iridates which have been studied via $\mu$SR including Eu$_{2}$Ir$_{2}$O$_{7}$ and Nd$_{2}$Ir$_{2}$O$_{7}$. While the bulk properties of Eu$_{2}$Ir$_{2}$O$_{7}$ strongly resemble those of Y$_{2}$Ir$_{2}$O$_{7}$ the local magnetic field extracted from $\mu$SR experiments indicate that the field is reduced by approximately 10$\%$ in Eu$_{2}$Ir$_{2}$O$_{7}$ compared to Y$_{2}$Ir$_{2}$O$_{7}$\cite{Zhao}. The PDFs' $f(\nu/\mu)$ and $g(\mu|\nu)$ for Eu$_{2}$Ir$_{2}$O$_{7}$ have been calculated from literature values\cite{Matsuhira, Zhao} with $g(\mu|\nu)$ shown along with the corresponding distribution for Y$_{2}$Ir$_{2}$O$_{7}$ in Fig. 4(a)-(c). As expected, these distributions are quite similar with Eu$_{2}$Ir$_{2}$O$_{7}$ simply shifted to a slightly reduced magnetic moment relative to Y$_{2}$Ir$_{2}$O$_{7}$.  To determine the extent to which structural differences influence these distributions, $f(\nu|\mu)$ has been calculated as a function of the lattice parameter \textit{a} and the oxygen parameter \textit{x}/a using the idealized $\mu$ = $\frac{1}{3}$ $\mu_{B}$/Ir and the $\psi_{1}$ configuration. The peak value of this distribution is shown in the insert in Fig 4(a), where \textit{a} is held fixed at 10.1669 {\AA} for varying \textit{x/a} and \textit{x/a}  = 0.34 for varying \textit{a}. It is clear that an increase in \textit{a} leads to a reduction of the maximum precession frequency at the muon stopping site, while the opposite is true for \textit{x/a}. This analysis shows that perturbations of the muon stopping site may in fact account for more than half of the observed difference in the precession frequencies, indicating that many such compounds likely share a similar underlying magnetic and electronic structure of the $Ir^{4+}$ sublattice.

\begin{figure}
\includegraphics[width=3.5in]{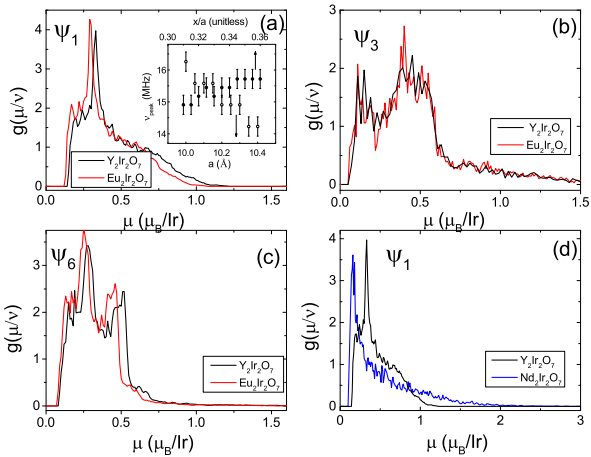}
\caption{(a) PDFs' $g(\mu|\nu)$ for Y$_{2}$Ir$_{2}$O$_{7}$ and Eu$_{2}$Ir$_{2}$O$_{7}$ for the $\psi_{1}$ configuration, the inset shows the variation of the peak in $f(\nu | \mu)$ calculated for an ideal $J =\frac{1}{2}$  moment as a function of a and x/a with a = 10.1669 {\AA} for variable \textit{x/a} and \textit{x/a} = 0.335 for variable \textit{a}; the error bars are given by the full width at half max of the peak. (b) and (c) show the $g(\mu|\nu)$ for Y$_{2}$Ir$_{2}$O$_{7}$ and Eu$_{2}$Ir$_{2}$O$_{7}$ in the $\psi_{3}$ˆ and $\psi_{6}$ˆ  configurations, respectively. (d) $g(\mu|\nu)$ calculated for Nd$_{2}$Ir$_{2}$O$_{7}$ where either the Ir or Nd lattice is ordered with the $\psi_{1}$  configuration}
\label{Figure4}
\end{figure}

While $\mu$SR measurements of Nd$_{2}$Ir$_{2}$O$_{7}$ clearly show long range magnetic order on the Ir$^{4+}$ sublattice\cite{DisselerNd227, Nd227New}, it has also been suggested that the Nd$^{3+}$ sublattice orders in a similar fashion\cite{Tomiyasu, Nd227New} with and ordered moment of approximately 2.5 $\mu_{B}$ \cite{Tomiyasu}. The Bayes' theorem approach has been applied to this system as well, using $\nu$ = 8.9 MHz as measured in Ref. \cite{DisselerNd227} and taking either the Ir or Nd sublattice to have the $\psi_{1}$ structure. This results in the $g(\mu|\nu)$ for each sublattice shown in Fig. 4(d). From this, it is clear that there is a vanishing probability that the Nd$^{3+}$ sublattice has an ordered moment near this expected value, and a peak near 0.15 $\mu_{B}$/Nd is observed instead. On the other hand, the $g(\mu|\nu)$ calculated for the Ir sublattice shows a sizable probability over a range consistent with that found for Y$_{2}$Ir$_{2}$O$_{7}$ and Eu$_{2}$Ir$_{2}$O$_{7}$, albeit with an slightly increased weight at lower values of $\mu$. Furthermore, a fully ordered Nd$^{3+}$ sublattice in this configuration would generate a spontaneous muon precession frequency near the proposed stopping sties of 75  $\pm$ 5 MHz, over an order of magnitude larger than that actually observed. These results show that the magnetic structure of the Nd$^{3+}$ sublattice cannot be described by this $\psi_{1}$ structure, and casts doubt on whether any magnetic structure as described by this same manifold may be realized on the Nd sublattice in this system.

The $\psi_{1}$ magnetic structure of Y$_{2}$Ir$_{2}$O$_{7}$ and Eu$_{2}$Ir$_{2}$O$_{7}$ determined here preserves the inversion symmetry of the lattice necessary to form the Weyl semi-metal state\cite{Wan,WK_PRB}. As this analysis has demonstrated, the Ir$^{4+}$ sublattice consists of ordered local moments near the full value expected from the $J_{eff} = \frac{1}{2}$ state, indicating that the \textit{5d} electrons are fully localized in these compounds; this would in turn suggest that these materials are best described by a fully gapped antiferromagnetic ground state rather than a semi-metallic one\cite{WK2013}. This is consistent with recent transport measurements of these systems\cite{Disseler2012,EuTransport}, and may be useful in placing restrictions on the values of the various hopping parameters used in calculations of band structures and phase diagrams\cite{Pesin, Yang_Ran,WK2013, Yang}. Furthermore, the apparent robustness of this structure suggests that similar ordering occurs on the Ir$^{4+}$ lattice in the weakly-metallic Nd$_{2}$Ir$_{2}$O$_{7}$\cite{DisselerNd227} which may contain a smaller Ir$^{4+}$ moment, and therefore may be more tenable hosts of such exotic phenomena. The lack of similar magnetic structures on the Nd$^{3+}$ sublattice supports other evidence for more complex ground state in this system\cite{DisselerNd227, Disseler2013, MatsuhiraGMR}, which may stem from an enhanced role of the \textit{f-d} exchange interactions similar to that in metallic Pr$_{2}$Ir$_{2}$O$_{7}$, \cite{Lee, Flint}.

In conclusion, both probabilistic modeling of the internal magnetic fields and \textit{ab initio} calculations of the muon stopping site have been used together for the first time here to determine the magnetic structure in different members of the pyrochlore iridate family using experimental data for the muon zero-field precession and lattice parameters. Specifically, it has been shown that the only magnetic structure consistent with the experimental observations is the $\psi_{1}$ and that the electronic state of the Ir$^{4+}$ is best described by the $J_{eff} = \frac{1}{2}$ state. These results have important consequences regarding the existence of exotic ground states in this class of materials by first demonstrating that the necessary symmetries are in fact present in the ordered state of these materials to support states such as the Weyl semi-metal, and secondly by further narrowing the range of compounds over which such states may be present in future studies.

\begin{acknowledgments}
The author would like to thank M. J. Graf and Stephen Wilson at Boston College and Stephen Blundell at Oxford University for helpful discussions during the preparation of the manuscript.

\end{acknowledgments}

\bibliography{BayesRefs}

\end{document}